# Dwelling Quietly in the Rich Club: Brain Network Determinants of Slow Cortical Fluctuations

Leonardo L. Gollo[1], Andrew Zalesky[2,3], R. Matthew Hutchison[4], Martijn van den Heuvel[5], Michael Breakspear[1,6]

[1] Systems Neuroscience Group, QIMR Berghofer, Brisbane, Queensland, Australia,

[2] Melbourne Neuropscyhaitry Centre and Melbourne Health, The University of Melbourne, Australia,

[3] Melbourne School of Engineering, The University of Melbourne, Australia,

[4] Center for Brain Science, Harvard University, Cambridge, MA, USA,

[5] Brain Center Rudolf Magnus, University Medical Center, Utrecht, Netherlands.

[6] Metro North Mental Health Service, Herston, QLD, Australia.

**Abstract:**

For more than a century, cerebral cartography has been driven by investigations of structural and morphological properties of the brain across spatial scales and the temporal/functional phenomena that emerge from these underlying features. The next era of brain mapping will be driven by studies that consider both of these components of brain organization simultaneously – elucidating their interactions and dependencies. Using this guiding principle, we explored the origin of slowly fluctuating patterns of synchronization within the topological core of brain regions known as the rich club, implicated in the regulation of mood and introspection. We find that a constellation of densely interconnected regions that constitute the rich club (including the anterior insula, amygdala, and precuneus) play a central role in promoting a stable, dynamical core of spontaneous activity in the primate cortex. The slow time scales are well matched to the regulation of internal visceral states, corresponding to the somatic correlates of mood and anxiety. In contrast, the topology of the surrounding "feeder" cortical regions show unstable, rapidly fluctuating dynamics likely crucial for fast perceptual processes. We discuss these findings in relation to psychiatric disorders and the future of connectomics.

## 1. Introduction: Looking Forward

Cerebral cartography has a long and storied past, founded on detailed examinations of structural and morphological properties to delineate regional boundaries - most famously, Brodmann's cytoarchitectonic maps (1) [for a recent historical perspective, see (2)]. These efforts continue in earnest today through labeling of functional subdivisions on the basis of receptor maps (3, 4) and, more recently, patterns of functional and structural connectivity (5-10). Although such maps become increasingly sophisticated and multimodal, they remain an inherently static snapshot of brain architecture, like the road map of a city. But brains may indeed be like cities, whose activities ebb and flow through its traffic network at different times of the day in ways that may not be immediately apparent from a map. With a map alone, the cyclic rush-hour of a metropolis and the overall change in travel direction (into and then out of the city) is not apparent. Moreover, roads co-evolve with the functional requirements of a growing city - constrained by multiple competing constraints of distance, time and history - such that structure and function are indelibly bound. In the same sense, having a detailed map of the cortex, even at the neuronal level, will not reveal the rich repertoire of dynamics that emerge from the underlying structure. It will also not be sufficient to use electrophysiological or neuroimaging approaches to map the resulting time-varying functional properties, even if these are eventually brought into the physical space of the white matter connections. There is an ongoing interplay at multiple temporal and spatial scales forming a chronoarchitecture (11, 12), requiring the two to be considered simultaneously.

Like a city built upon moments of frenetic day time activity, with its daily cycle of rush hours, that groans relentlessly through its winter months and relaxes in the summer, the brain is a complex, dynamical system composed of many time scales. And like a city that rushes at pace in its centre but not in its suburbs, the time scales of the brain likely vary across spatial scales and anatomical regions. At present, when mapping the dynamics of the brain is becoming as legitimate as mapping its receptors and connections. This enquiry can benefit substantially from mathematics, where topology and geometry have been used to map the nature of dynamical systems since the pioneering work of Poincare (13). Here, we consider the origin of nested time scales in the brain and look prospectively toward their relationship with the time scales of cognition and emotion.

*1.1 Slow Rhythms and Hierarchical Time Scales*

The rapid "chattering" of neural cell activity originating from ion channel opening, cell spiking, and bursting varies across fast time scales, from microseconds to hundreds of milliseconds. These processes are well matched to the fast temporal and spatial timescales of natural scene perception (14)- i.e. there is an approximately 1:1 relationship between the time scales of perception [e.g. (15)] and the time scales of fast neuronal dynamics in sensory systems: Indeed much of neurobiology research is predicated upon this assumption, as captured succinctly by the proposal that the complexity of neural systems are matched to those of scene statistics (11, 16).

However, many important mental processes unfold on much slower time scales than the fast chattering of sensory systems. The subtle fluctuations in affect that occur during human social exchange occur on the scale of seconds. Underlying mood fluctuates even more slowly (17). Human movement scales continually across a broad spectrum of time scales, from seconds through to the diurnal variations imposed by sleep (18). Human movement patterns show a very broad power spectrum. They additionally conform to a nested hierarchy such that fast behaviours are nested within slower ones (19). Physiological states, such as respiratory and cardiac related fluctuations occur on intermediate time scales. Likewise, the spatial and temporal statistics of the natural world have a 1/$f$-like spectrum (20), meaning that most of the power in the natural world is in the very low spatial frequencies and slow time scales. If the human brain is to find a match with the slow time scales of these scene statistics, then slowly fluctuating brain states are required. What are the neural correlates that are matched to these slow time scales?

Classic neurobiological accounts of these slower times scales invoke synaptic processes (21). These begin with short-term depression, short-term potentiation, and spike-time dependent plasticity (STDP) (22) for the time scales immediately slower than neural spiking, moving to morphological and transcriptional processes for slower time scales. For example, introducing STDP into a model of an ensemble of spiking neurons disrupts high frequency oscillations, yielding 1/*f*-like non-oscillatory temporal activity (23) and broad spectrum, avalanche-like spatial processes (24). Likewise, hierarchical theories of long-term learning, behaviour, and memory have found their counterpart in hierarchical models of reinforcement-based synaptic updates (25) and synaptic processes (26).

There is now considerable evidence that cortical anatomy recapitulates the temporal hierarchy that is inherent in the dynamics of environmental states and human behaviour (27). Classically, this anatomical arrangement appears to be primarily based upon a caudal-rostral (fast-slow) gradient. Prefrontal regions map the slow aspects of natural statistics (27) and human behaviour (19, 28-30), mirroring the hierarchical arrangement of increasingly complex cognitive functions (31, 32). A caudal-rostral gradient has even been proposed as ordering hierarchical relationships at finer spatial scales, such as motor control systems within the frontal lobe (32, 33). Whilst neurons might spike very quickly, slower fluctuations in the rates of these spikes may be the key to this hierarchy of slower time scales. A recent compilation of the timescales of intrinsic fluctuations in spiking activity in primates reveals a hierarchical ordering, with sensory and prefrontal areas exhibiting shorter and longer timescales, respectively (34, 35). Using movie streams cut into segments of different lengths, Hasson *et al.* likewise revealed a caudal-rostral hierarchy of cortical time scales (36, 37) that mirror the conceptual frameworks of Fuster and Mesulam (30, 31).

*1.2 The Role of Computational Models*

Computational models of brain dynamics are of considerable interest in systems neuroscience because of their potential to compliment the descriptive approach of empirical research with generative models, mirroring the importance of models in the traditional "hard branches" of physics (38-40). Models of large-scale neuronal activity have been applied to variety of healthy and pathological phenomena, including resting state fluctuations (41, 42), perception (43) and seizures (44, 45). What theoretical frameworks can address the origin of slow time scales in brain and behaviour? In addressing this question, we seek a framework that goes beyond the broad time scales that can be obtained through equipping models with a "reservoir" of broad time scales (34, 46), towards one that integrates (or nests) the spectrum of time scales in a natural hierarchy, similar to that proposed for the arrangement of nested brain rhythms (47-49).

One simple approach, drawing on the treatment of time scales in physical systems (50) is to construct a hierarchy of dynamical systems: Consider an ensemble of *N* physical systems where the states of each system (firing rates, membrane potentials, etc.) are represented by a vector [$\mathbf{X}_i$], where *i*=1, 2, … , *N*. In the autonomous (uncoupled) case, the dynamics of the *i*-th system are given by a differential equation,

$$T_i \frac{dX_i}{dt} = F(X_i), \quad (1)$$

where *F* is an arbitrary nonlinear function that embodies the properties of each system and the *time scale multiplier* $T_i$ imbues each system with its own characteristic time scale. By specifying these constants to be drawn from a geometric series, $T_i = \tau \times T_{i-1}$ the system of Eq. (1) is naturally endowed with a (power law) hierarchy of time scales. That is, the temporal relationship between neighbouring subsystems is the same across the entire ensemble: the ratio between the time scales of any two subsystems *j* and *k* is given by $T_j/T_k = \tau^{(j-k)}$.

These systems can then be nested, or coupled, by driving each system by its faster neighbour,

$$T_i \frac{dX_i}{dt} = F(X_i) + c \cdot F(X_{i-1}), \quad (2)$$

such as the fast fluctuations in primary sensory cortex driving the states of a slower association region. The scalar *c* is a coupling parameter that modulates the strength of this coupling. The key to this arrangement is that autonomous nonlinear systems *F* (both limit cycles and chaos) classically show a peak frequency as well as strong harmonics (2:1, 3:1, 3:2, …) and subharmonics (1:2, 2:3) – a feature that has indeed been demonstrated in primary somatosensory (51) and visual (52) cortex. If the time scale multiplier is close to an integer or simple fraction ($\tau$ = 3/2, 2, 3, …) then the energy at one of these peaks can entrain the subordinate's system fundamental oscillation. In this way, the faster systems form a progressive series, adjusting the phase and peak frequency of the slower systems toward the harmonics of their own (Figure 1).

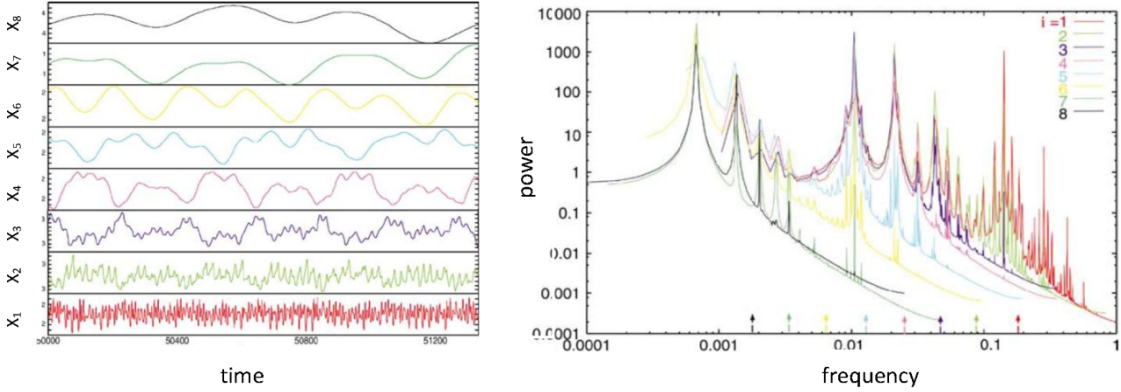

*Figure 1:* Time series and power spectra of nested time-scales hierarchy in an ensemble of coupled nonlinear oscillators. Each system has a unique colour. (Left) Time series from slow (top) to fast (bottom). (Right) Corresponding power spectra from fast (right) to slow (left). The fundamental and harmonic frequencies of the slow systems (left hand side of the spectra) overlap the harmonics and subharmonics of the faster, driving systems. Arrows show natural (uncoupled) frequencies of each system. Note that the intrinsic time scales are uniformly positioned along the logarithmic axis. Adapted from Figure 1 of (50).

Two computational accounts of time scale hierarchies in the brain leverage this framework. In (53), a multiscale wavelet framework was employed so that the ensuing temporal hierarchy was embedded in a corresponding spatial hierarchy: That is, fast microscopic elements drove slow, coarse-grained neural systems. Although in principle, this approach could be adapted to accord to a caudal-rostral hierarchy, the objectives of (53) were rather to model interactions across spatial and temporal scales *within* any particular cortical region – such as the influence of single unit spiking activity on collective mean field macroscopic oscillations [see also (54)]. In contrast, Kiebel & Friston (27) did map the time scale hierarchy of Eq. (1) onto the rostral-caudal hierarchy of cortical anatomy. However, the coupling between time scale systems in (27) was not achieved in the manner of direct coupling, i.e. Eq. (2), but rather through a hierarchy of metastable heteroclinic cycles. That is, the autonomous dynamics of Eq. (1) were constructed so that the (fast) dynamics progressed around a cycle of fixed point saddles. The states of each successively slower dynamical system were then coupled to these emergent cycles.

Whilst both of these accounts address the nesting of neuronal timescales, this was largely achieved through construction. That is, the systems reproduce the temporal hierarchies seen in the brain largely because they are constructed to do so ([26](#), [27](#), [53](#)). The treatment of time scales is largely heuristic in previous studies because it is imposed upon the ensemble of systems through the use of the time scale multiplier. There are several candidate mechanisms underlying these heuristic treatments of time scale hierarchies. The time scales embodied by varying synaptic processes (fast AMPA, slower NMDA, slow kainate, etc.) are an obvious choice. Such synaptic and cellular substrates of long-term behaviour and memory are the focus of a rich body of research that we will not review here. Instead we ask whether the topological arrangement of neural populations into circuits and networks provide a scaffold on which slow and stable dynamics emerge? Hence, we will neglect morphological changes in neurons and focus only on temporal hierarchies as they arise through the arrangement of neuronal interactions in the brain.

*1.3 Objectives of the Present Study.*

Analyses of anatomical networks of brain systems have repeatedly emphasised their inherent hierarchical structure ([31](#), [55-57](#)) Recent studies in particular, have highlighted an inner core of the brain, namely a densely interconnected constellation of hub regions called the *rich club* in human ([58](#), [59](#)), macaque ([60](#), [61](#)) and even the *C. elegans* brain ([62](#)). On the periphery of this core is a second level of *feeder* regions that are not hubs although they do connect directly to the rich club. Outside of these levels are the *peripheral nodes* which are sparsely connected to other nodes and do not connect to the rich-club regions. Is it possible that this inherent core-periphery structural hierarchy of regions confers a dynamic hierarchy? That is, could a spectrum of dynamics emerge through ensemble interactions on a structural hierarchy and not from biochemical factors? Intriguingly, many rich club hubs - such as the anterior insula and dorsal anterior cingulate cortex ([60](#)) - overlap with those associated with slow cognitive processes such as mood ([63](#)).

The objectives of the present study were to study the dynamics that emerge from the interaction of identical dynamical elements coupled via the structural connectome of primate cortex. That is, we do not impose a hierarchy of time scales, but rather start with an ensemble of neural mass oscillators which all have the same characteristic time scales. We then partition their dynamics according to the rich club hierarchy to find an emerging dynamical hierarchy. The paper is structured as follows: In the next section, we describe our model of neuronal dynamics on the anatomical connectome of the macaque cortex. We first review the presence and nature of the rich club in these structural connectivity data. We then analyse the dynamics that emerge, showing that the dynamics in rich club nodes are more stable, with greater temporal persistence and dynamic stability than peripheral regions. We also show how rich club nodes coordinate the dynamics of their neighbours through synchrony, thus promoting integrated dynamics. We then characterize the main factors in shaping the network behaviour: hubs, connectivity, motifs and time delays. In the final section, we contextualise these findings in the light of recent research areas, drawing out possible synergies between the morphological features of rich club and the time scales of emergent dynamics. Finally, we speculate upon the role of slow time scales in rich club nodes - particularly the anterior insula - and the relatively slow time scales of interospection and affect.

## 2. Dwelling in the rich club

*2.1 Model construction: Structure and dynamics*

We modelled neuronal dynamics unfolding on an anatomical network reconstructed from primate cortex. For this, we used a fine-grained version of the collation of anatomical connectivity data for the macaque (CoCoMac), (64). These data represent the right hemisphere of the Macaque cortex as a network of 242 nodes, with connections derived from published tracing studies (60). As previously reported (60), this network possesses a modular structure with 5 distinct communities (Figure 2a,b) as well as a clearly defined rich club core (Figure 2c). Ordering the nodes into their five modules and plotting their degree shows that rich club hubs are distributed across each of the network's modules (Figure 2e), hence acting as network hubs both at the whole network level, in addition to linking individual communities. Note that the sizes (number of nodes) of each module differ considerably and the number of rich club nodes within each module is not proportional to the size of the module.

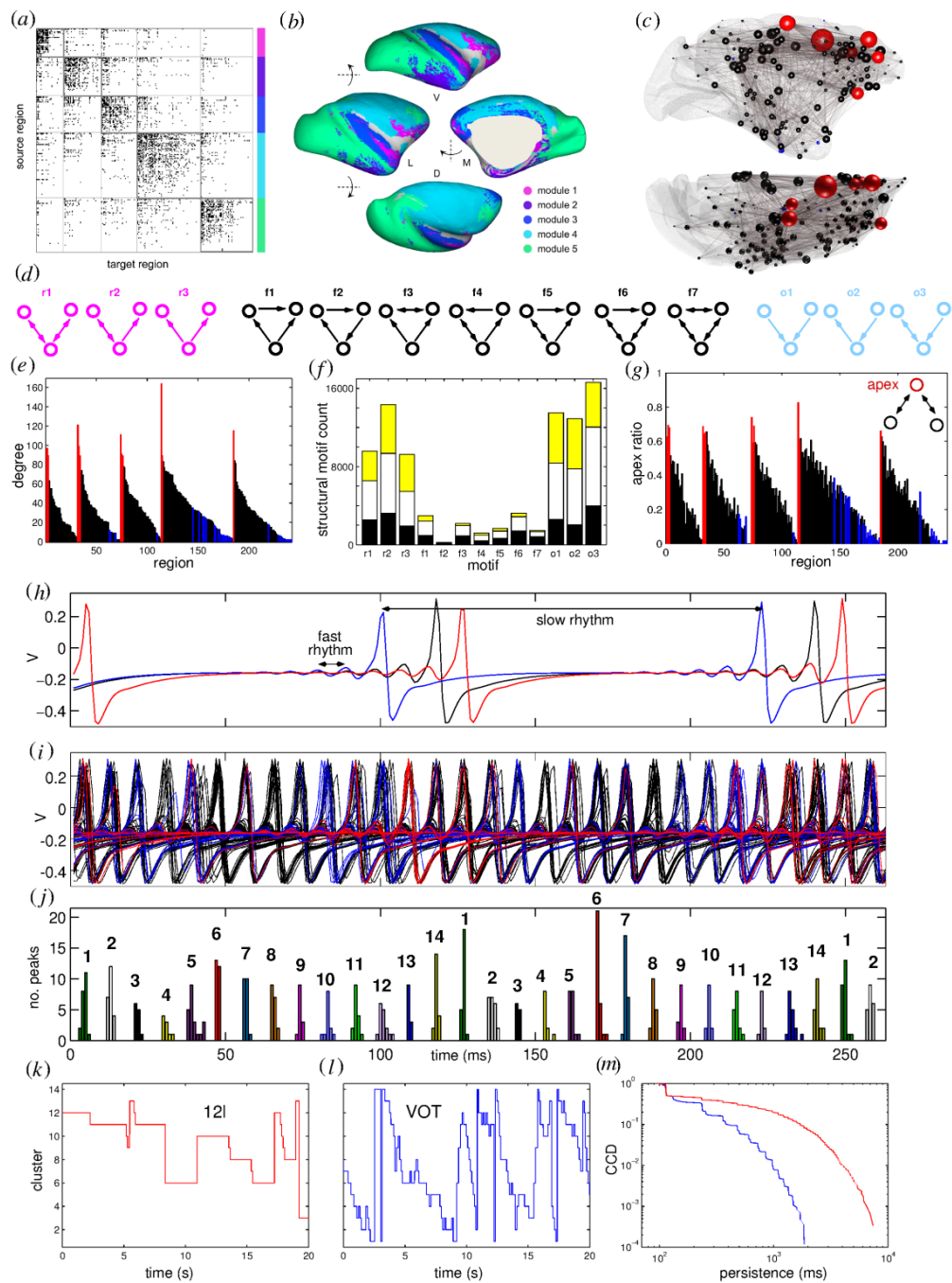

*Figure 2. Structure and dynamics of simulated model. (a) Binary connectivity matrix with 242 regions, 4090 directed connections. (b) Inflated right hemisphere of macaque brain; Panels (a) and (b) are adapted from (60) and show that the network decomposes into 5 modules. (c)*

*Anatomical representation of the network, with nodes colour-coded by club membership [red = rich club (R); black = feeders (F); blue = periphery (P)] and degree denoted by size. (d) Collection of all 3-node motifs. The 13 motifs are divided into three families: Resonant (r1, r2, r3); Frustrated (f1-f7); and Other (o1, o2, o3). (e) Degree of nodes, ordered into their modules and colour-coded according to (c). (f) Structural motif count; black bars represent proportion of motifs with only intra-cluster connections, yellow bars motifs with only inter-cluster connections, and white bars motifs with both types of connections. (g) Proportion of counts in which each node occupies the apex position of 3-node motifs, including all types of motifs. Rich-club members are shown in red. (h) Illustrative time traces of three neural mass models showing its two oscillatory rhythms. (i) Time traces corresponding to all nodes. (j) Functional clusters; the set of phase-locked regions are divided into 14 independent clusters based on the co-occurrence of their slow rhythm peaks (the time series of panels i and h are temporally aligned). Exemplar functional cluster dynamics for 12l, a rich-club region (k), and for VOT, a peripheral region (l). (m) Complementary cumulative distribution of 12l (red) and VOT (blue).*

The "building blocks" of this network - namely the small network motifs (65) - play an important role in shaping neuronal dynamics, particularly in the presence of time delays (66, 67). Structurally they show an uneven arrangement across the network (60) that we briefly re-visit. Based on our previous analyses (66, 67), we divide the 3-node motifs into three families according to their dynamical properties (figure 2d): Resonant motifs (r1-3) are open motifs with edges directed away from a central node and favour zero-lag (isochronous) synchronization of the outer nodes (66); frustrated motifs (f1-7) are closed motifs which promote metastable dynamics (67); other motifs (o1-3) do not easily fit into these families and show divergent dynamics. Frustrated motifs appear less frequently in the connectome than resonant and other motifs (figure 2f). As evident in the colour coding of figure 2f (specifically, the lack of yellow), frustrated motifs also tend to connect nodes within modules: Only 14% of frustrated motifs connect nodes in different clusters, in contrast to 35% for all other motifs. All other motifs comprise connections within and between modules. Rich nodes (60) typically occupy an apex position on 3-node motifs (figure 2g) and mediate most minimum path length routes through the connectome (62, 68, 69). We return to the role of these motifs in shaping network dynamics later.

To study the collective dynamics and time scales unfolding on this network, we defined neuronal dynamics on each node then connected them via the structural edges. Nodal dynamics were obtained by simulating a conductance-based neural mass model (70, 71). The dynamics of each neural mass is described by three coupled non-linear differential equations for the subpopulation of excitatory, inhibitory neurons, and the dynamics of slow potassium channels (see Appendix A1). Without the inhibitory population, the system is essentially a population of simple spiking neurons: The presence of the inhibitory neurons, which have a relative slow response, leads to chaotic dynamics in the autonomous single node (uncoupled) case (70). Unless otherwise stated, regions were coupled with a relatively weak coupling strength, c=0.01, and an axonal delay of 15 ms although the results are robust across a range of these values. All other parameters are shown in Appendix A1 as per previous work (66), and the simulation details are described in Appendix A2.

*2.2 The rich club as a stable dynamic core*

With this set of parameters the neural masses oscillate with two time scales: a fast fluctuation of ~110 Hz superimposed on a slow oscillation of ~8 Hz (figure 2h). With the weak inter-node coupling, regions synchronise at the fast time scale. Synchronization at the slower time scales occurs in clusters so that several of these regions peak in unison at the high amplitude peak (characteristic of the slow oscillatory rhythm; figure 2i). We named each assembly of regions peaking together a "functional cluster". As there are 14 oscillations of the fast rhythm within each period of the slow rhythm, there are 14 functional clusters, which we label and dynamically track. In figure 2j we provide an illustration of the number of regions peaking within time bins of 1 ms, and enumerate each of these 14 non-overlapping functional clusters. If the interval between two consecutive high amplitude peaks for any given node is similar to the average inter-peak interval (IPI) then that peak will remain in the same functional cluster. However, regions often jump between functional clusters either by advancing or delaying their IPI with respect to the average. This is evident in the unrelenting shift in the number of nodes within each cluster as the network evolves. Rich-club regions, like the prefrontal region 12l shown in figure 2k, reside with greater stability[1] within their functional clusters, hence exhibiting a more persistent behaviour when compared to the rest of the network. A rich-club member, for example, can remain in the same functional cluster for several seconds. In contrast, peripheral nodes, such as the ventral occipitotemporal area VOT illustrated in figure 2l, are less stable and exhibit a less persistent behaviour, frequently jumping back and forwards between functional clusters. Figure 2m shows the complementary cumulative distribution, which quantifies the persistence time of the two regions (a rich node, 12l and a peripheral area, VOT) within a given cluster before a cluster transition takes place. Note the fatter right hand tail of the rich club node (red) – denoting cluster dwell times of almost an order of magnitude longer than the peripheral node. These cluster transitions do not occur uniformly in time, but rather in intermittent periods of high instability.

To characterise the dynamics of each node, we used a simple measure of irregularity, namely the difference between each node's IPI and the ensemble mean IPI. This measure of irregularity, which captures how often nodes jump between functional clusters, differs substantially among nodes and shows a dependence on the nodes' club. For example, the average irregularity for all rich-club nodes is 3.9 ms (for each long cycle) compared to 6.3 ms for peripheral nodes. A general picture of the spatial distribution of this measure is shown in figure 3a, with specific values for each node presented in figure 3c. IPIs of the peripheral (/rich) nodes are more (/less) irregular than the network average. That is, rich nodes stay far closer to the global network oscillation than peripheral nodes. To measure local dynamic stability, we also quantified the dynamics of pairs of nodes by measuring how often each pair changed their phase relation per second. For each node, we then averaged this pair-wise value across all 241 (N-1) possible pairs (figure 3b,d). This pair-wise measure exhibits similar behaviour to the node-wise measure of irregularity, with rich (peripheral) nodes more (less) stable.

---

[1] The term "stability" does not refer here to the dynamics of each node (which are all chaotic), but rather to the "transverse stability" of pairs of nodes - that is, the stability of their tendency to synchronise ([72]). The stronger transverse stability of the rich club nodes to their neighbours underlies the less frequent phase re-adjustment and hence the longer dwell times. Formal analyses of transverse stability in this system are found in fig. 8 of ([70]) and fig. 9 of ([66]).

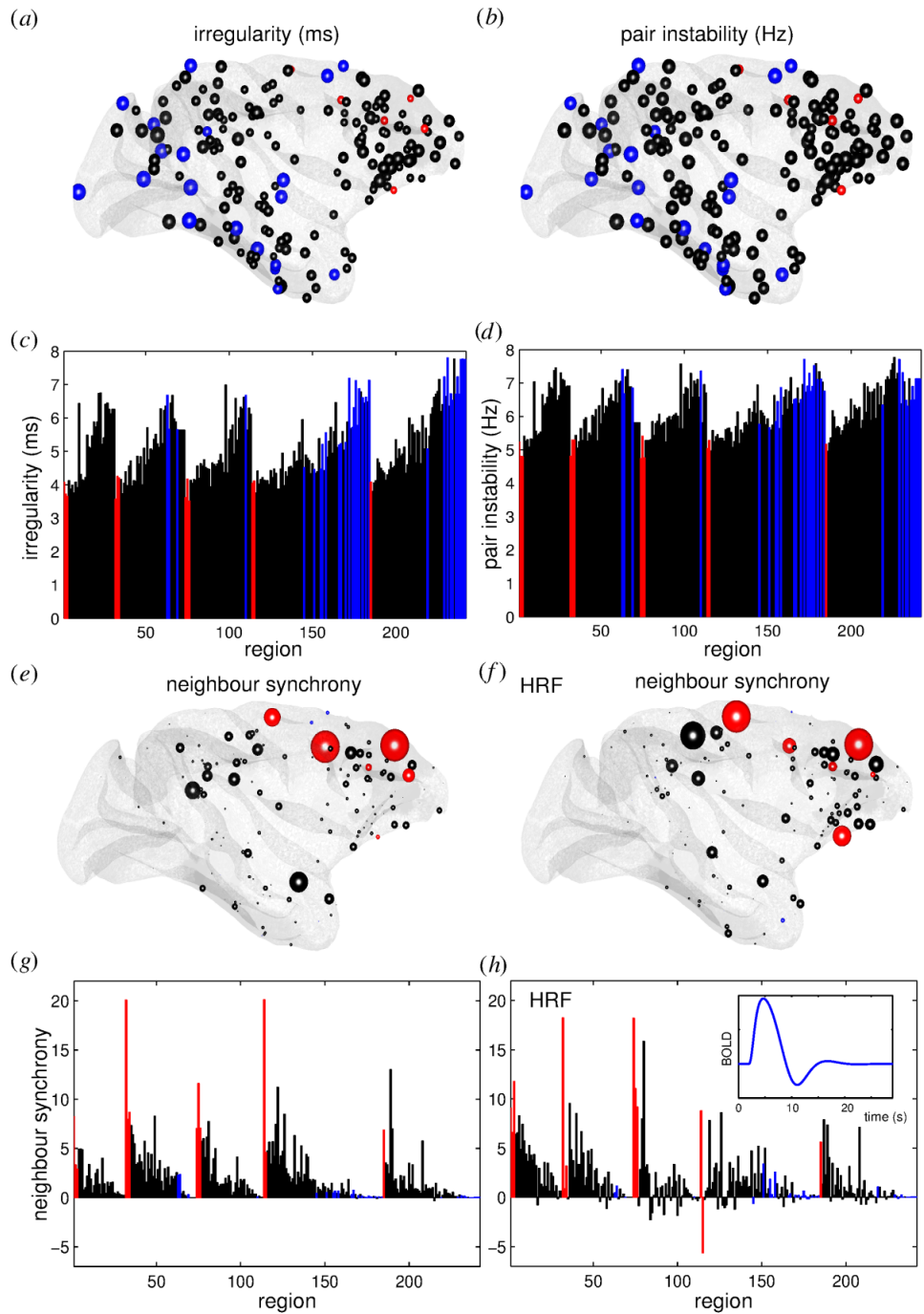

*Figure 3.* *Dynamical core. Irregularity of IPI; (a) representative network [of Fig 1 (c)] and (c) bar plot. (b), (d) Average number of changes in phase relation per second for all pairs that each node participates in. Average synchronization between pairs of neighbours of 3-node motifs for the membrane potential of the excitatory subpopulation V (e) and (g). Results represent an average over 20 trials of 90 seconds. (f) and (h) same as (e)*

*and (g) but for the hemodynamic response function (inset) in a long trial of 600 seconds.*

We next shifted our focus from measures of local dynamic stability to metrics of network integration. As an initial measure of local integration, we measured zero-lag synchronisation of each node's neighbours, averaged over all neighbour pairs and then normalised (see Appendix A3). Anatomically high-degree (rich) nodes play a prominent role in mediating local synchronisation between their neighbours, with values up to an order of magnitude higher than other nodes (figure 3 e.g.). Notably, this property persists after convolving neural activity with the hemodynamic response function, showing that rich nodes also promote neighbour synchrony at this much slower time scale (figure 3f,h). The preservation of the results across different time scales suggests that the role of the rich-club nodes in promoting neighbour synchrony may also be evident in resting-state fMRI data.

*2.3 Network determinants of slow, stable dynamics.*

The dynamics of rich-club nodes are thus regular, stable and catalyse zero-lag synchronisation between their neighbours across slow time scales. We next sought a deeper understanding of anatomical determinants of these findings by systematically varying the topology of the networks. To achieve this, we compared the dynamics of the CoCoMac network with six different randomised networks: A simple random network (preserving only nodes and edge numbers); a degree-preserving random (DPR) network; three randomised networks which eliminate bidirectional connections from each of the previously mentioned networks; and a random network with uniform in-degree. These benchmark networks allows us to discern which dynamic properties are due to the heterogeneous node degree, and which owe to higher order properties such as the motif composition, rich club structure and network symmetry (i.e. the role of bidirectional edges).

We first sought to quantify the overall dynamic heterogeneity across each network: That is, which factors cause some nodes to be more regular than others. This was achieved by simulating the dynamics and then measuring the coefficient of variation (CV, the ratio of the standard deviation to the mean) of dynamic irregularity across all nodes. A small CV denotes networks with homogeneous dynamics. Comparing the different network topologies (figure 4a), we find that the simple random networks, which have a more homogeneous degree also exhibit a more homogeneous irregularity. The class of random networks with completely uniform in-degree (open circle) has the most homogeneous irregularity. The degree-preserving random networks, however, have a similar dynamic heterogeneity to the original CoCoMac network. This suggests that the high node in-degree, which plays a crucial role in the stable, regular dynamics of the rich club, plays a crucial role in creating dynamic heterogeneity across the network. In other words, more complex dynamics arise in networks with more complex topological arrangements than random networks.

Interestingly, on this measure, the networks lacking bidirectional connections are similar to their counterparts. However, a comparison between the networks for the overall irregularity of the network (figure 4b) reveals the importance of bidirectional connections: Randomising edges to eliminate bidirectional connections (black symbols) reduces the overall irregularity of the network. Consistent with previous work (38, 73) conduction delays also play an important role in the irregularity of network dynamics, which is substantially reduced in the absence of delays [figure 4c, (please note the different scales on y-axes in figures 4b and 4c)].

Dynamic irregularity can be viewed as a proxy for metastable transitions in the network. A small value of irregularity means that nodes stay within their functional clusters with stable oscillations and with few metastable transitions. Metastability has been associated with frustration (67), which arises in the frustrated (closed) motifs in the presence of inter-node delays because there is no single pattern of pair-wise stable zero-lag synchrony (66). Since most of the frustrated motifs involve bidirectional connections, the elimination of bidirectional connections may affect the irregularity by reducing the overall frustration and hence metastability of the network. Eliminating delays is even more effective in reducing the irregularity (figure 4c) because it is a crucial requirement for metastable dynamics to arise in frustrated motifs (67).

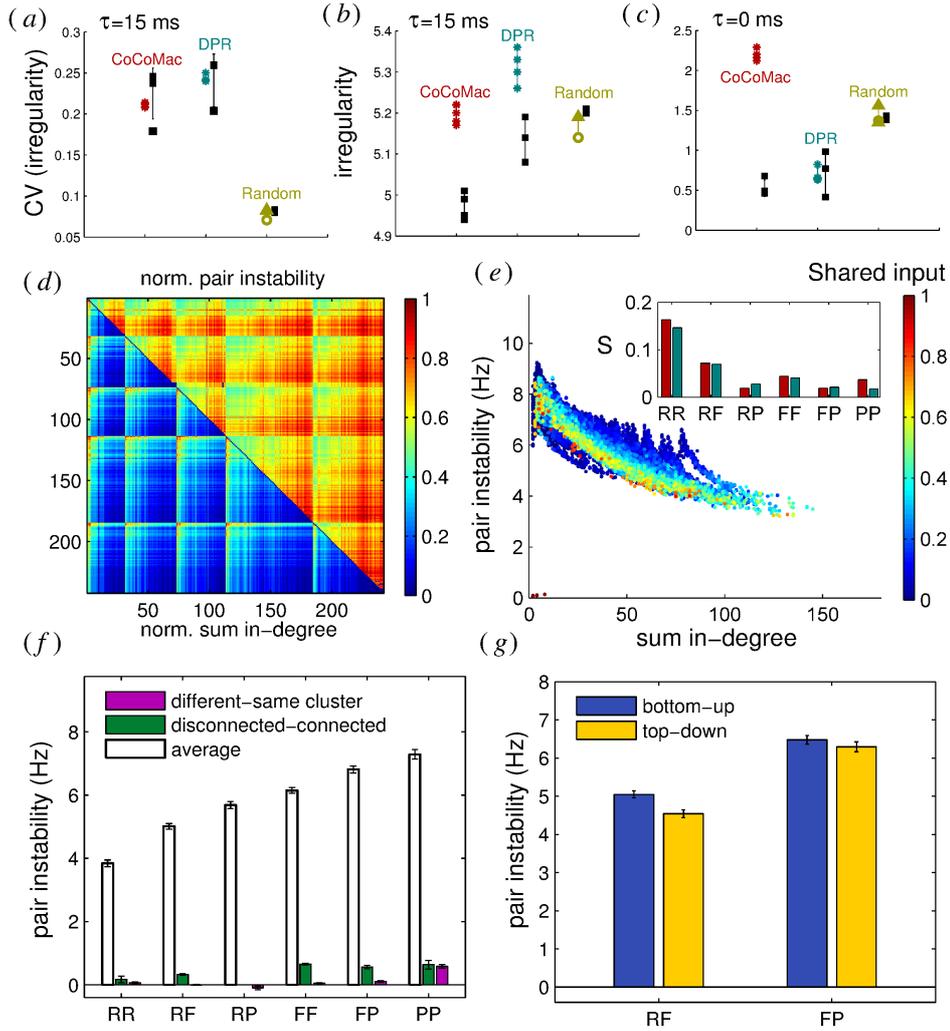

*Figure 4.* Role of topology in dynamic timescales. (a) Coefficient of variation of the irregularity, and (b) irregularity for the CoCoMac, degree-preserving random (DPR) network, random network, random network with homogeneous in-degree (open circle), and each of these network randomized to eliminate all bidirectional connections (black symbols next to each network). (c) Same as (b) but with no time delays. (d) Normalized average number of changes in phase relation (top) and normalized sum of in-degree for each pair of nodes (bottom). (e) Number of changes in phase relation per second versus sum in-degree. Colours depict representation of the proportion of shared input. Inset, Average proportion of shared input for the different type of node pairs for the CoCoMac and the DPR network. (f) Average number of changes in phase relation (white), difference between disconnected and connected pairs (green), and difference between pairs of different clusters and of same cluster (purple). (g) Average number of changes in phase relation for unidirectionally connected pairs from the
16

*second to the first (blue) and from the first to the second (yellow) type of nodes.*

The node in-degree has been previously shown to also play an important role in shaping the network stability in the presence of strong coupling (74). Here we find that node degree influences the pair instability as well as the dynamic irregularity. For a pair of nodes, is their total in-degree (i.e. the sum of each node's in-degree) the only factor influencing their synchrony? In figure 4d we see that the (in)stability of a node pair's synchrony is (anti)correlated with that pair's summed in-degree. This anti-correlation is also evident in a scatter plot of pair instability and total pair in-degree (figure 4e). In this case, most pairs belong to a single cloud, with the exception of a few pairs that have small in-degree (see lower left corner). To investigate these outliers, we defined a metric $S$ as the proportion of shared input of each pair dyad to their total in-degree (see colourbar). The outlier nodes have the maximum value of this metric (i.e. each node receives exactly the same input as the other) and almost complete synchrony. However, even a small mismatch in the pair's input substantially increases the pair instability and drives it back into the cloud. Aside from the outliers, pairs with most shared input occupy the centre of this cloud, which follows a nonlinear relationship between pair in-degree and pair stability. Furthermore, we find that the shared input is enriched in the CoCoMac network (red) comparing with the DPR network (blue), between pairs of rich and pairs of peripheral nodes (inset of figure 4d). This weaker shared input may explain the greater overall dynamic irregularity in the DPR network (figure 4b).

In addition to the proportion of shared input, the pair instability is also shaped by the type of nodes (rich, feeder, periphery) in the pair, the connectivity pattern between the pair, and the modularity of the network (figure 2d). Pairs of rich nodes are the most stable pairs whilst pairs of peripheral nodes are the most unstable (figure 4f; white bars); this is consistent with the anti-correlation between pair instability and the sum of their in-degrees. Connected pairs are more stable than disconnected ones (green bars), particularly for peripheral pairs. Modularity plays almost no role for rich pairs, but does play a small role for peripheral nodes (purple bars). The direction of unidirectional edges can also influence the pair instability (figure 4g). Connections from the core to the periphery ("top-down", yellow bars) are more effective at stabilising the dynamics of pairs (= lower pair instability) than connections from the periphery to the core ("bottom-up", blue bars). This is consistent with the salience of top-down influences in cortical hierarchies ([75]). Overall, these results also show that the dynamics of the periphery are more susceptible to basic network features like connectivity and modularity.

Rich-club regions are the most crucial regions in promoting neighbour synchrony and dynamical integration. A further structural contribution for this result comes from the fact that high-degree nodes are apices of a larger proportion of motifs that they participate in than low degree nodes (figure 2g). In addition to the proportion of apexes, the type of motif (figure 2d) for which a node is an apex also exerts a substantial influence in the fate of the synchronisation between neighbours. For example, an apex of motif o1, which only receives unidirectional connections from its neighbours, does not contribute to the synchronisation between such pair. On the contrary, an apex of motif r3 effectively contributes for the synchronisation between its neighbours. We next further pursue the role of motifs in network dynamics.

*2.4 Role of delays and network motifs.*

We previously found that in the presence of time delays the structure of isolated 3-node motifs influences local zero-lag synchrony ([66]). In particular, resonant motifs promote zero-lag synchrony between their outer pairs – that is, when an open motif has a bidirectional connection at its apex (either to its outer nodes, or to other motifs). In the CoCoMac network, 99.4% of r3 motifs have a bidirectional connection at the apex node. Rich club nodes are the apex of a large number of resonant motifs (red bars, figure 5a,b), but only few frustrated motifs. In contrast, frustrated motifs are over-represented as apices of feeder (black) and peripheral nodes (blue bars). The apices of other motifs are uniformly distributed across all three classes. How do these motif arrangements influence local synchrony? In the absence of delay, both resonant and frustrated motifs promote neighbour synchronisation. Hence, the neighbour synchrony in the absence of delay (white bars of figure 5c,d) is positive for most motif types and all families. However, in the presence of axonal delays (orange bars) resonant motifs boost the synchrony of their neighbours, whereas frustrated motifs are now associated with negative neighbour synchrony. As a control experiment, in the uncoupled case (black bars), neighbour synchrony is, as expected, almost zero for all motifs. The dynamical repertoire associated with isolated motifs hence survives immersion into the primate connectome and sculpts patterns of zero-lag synchrony. Moreover, the enrichment of resonant motifs with rich club apices appears to diminish local dynamical frustration, promoting a zone of stable, synchronous dynamics.

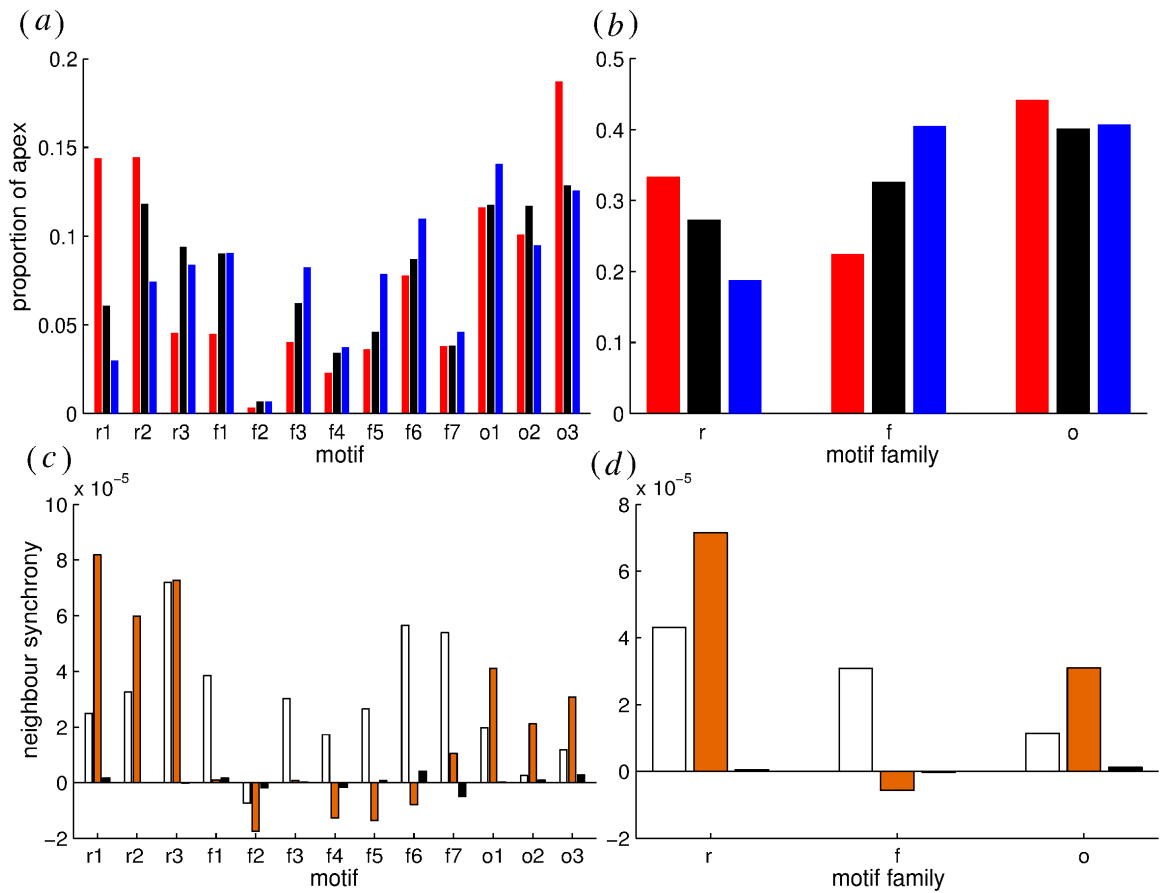

*Figure 5.* *Role of network motifs in catalysing local synchronization. Proportion of counts in which each motif (a) and each family of motifs (b) contributes to the total number of apex counts for rich (red), feeders (black) and peripheral nodes (blue). Average neighbour synchronization per motifs (c) and per family of motifs (d) for the cases of no-delayed connections (white), delayed connections (orange) and disconnected networks (black).*

## 3. Discussion: Chronoarchitecture of the brain

*3.1. Rich club nodes: Masters or slaves of network dynamics?*

Although a multitude of neuronal, synaptic and morphological processes contribute to the hierarchy of time scales in the brain, here we focused on the role of network topology. Our analyses suggest that the high in-degree of rich club nodes plays a central role in promoting a stable, dynamical core of spontaneous activity in the primate cortex and highly unstable dynamical transitions in the periphery. The arrangement of motifs and the local dynamics they promote plays an additional, sculpting role – with the rich club benefitting from its disproportional position at the apex position of synchrony-promoting resonant motifs. In contrast, closed motifs - which support fast, metastable interactions - are found throughout feeder and peripheral regions, which jump erratically between functional clusters. This fits with our proposal that the time scales of peripheral (sensory) regions should be matched to the faster perceptual processes locked to the external milieu, whereas core (multimodal and limbic) regions should possess greater and slower dynamic stability. Our findings add to the small, but intriguing body of empirical (76), computational (77) and theoretical (78-80) work on the presence a global dynamic core of the primate brain by showing how this feature of chronoarchitecture can arise through network topology alone.

Rich club stability arises largely through the high in-degree of these regions. Intriguingly, as a metaphor to social networks (81, 82), rich-club members could be seen to be "slaves" of their own power. The richness of these brain regions comes from their large in-degree. However, for this very reason, their dynamics are consequently more regular because larger in-degree guarantees limited irregularity. In other words, the autonomous dynamics of connectome hubs are largely enslaved to the strong, rhythmic output of the entire connectome. This may relate closely to their role in sampling the collective dynamics of cortex. Therefore, rich-club regions cannot behave very differently from what is "expected" in contrast to the peripheral nodes which have the most freedom to explore the dynamical landscape. Indeed, whilst the rich club leads to a stable core, it's existence may paradoxically increase the dynamic flexibility in low-degree peripheral regions, hence increasing the brain's overall functional diversity (83). That is, the presence of a stable rich club may release peripheral nodes from synchronisation. Such a paradoxical process may be just as important to adaptive cognitive function as a stable dynamical core, because unstable dynamical transients may play a crucial role in sensitive perceptual processes in the face of ambiguous stimuli (84-86).

*3.2. Anatomy, time scales, morphology*

In accordance with the caudal-rostral hierarchical time scales, rich-club regions are generally located in more anterior regions (fig. 2c). In the macaque connectome that we analysed, anterior rich club regions are located in prefrontal, para-cingulate and anterior cingulate cortex: Less rostral exceptions include the parietal areas 7b and LIP and temporal areas TH and TF. Hence, more stable and slowly evolving dynamics are generally shifted rostrally. However, our dynamics-based approach suggests that a caudal-rostral mapping of the temporal hierarchy should be treated as a "first approximation" to the problem of neural time scales. A "post-connectome" view suggests that the anatomical mapping of the temporal hierarchy may be better described as a periphery-rich-core axis. Human connectomes reconstructed from non-invasive diffusion-based tractography suggest prominent rich club regions in posterior midline cortex and around bilateral inferior frontal gyrus (58): Hence, a rostro-caudal axis may be less important in the primate cortex than a centre-periphery axis.

Although we focus upon spontaneous, endogenous dynamics in the primate cortex, other work (87) has advanced similar principles operating over task-relevant cortical systems during execution of specific tasks. This work converges with our own, in the sense that a stable dynamic core was associated with dense (functional) connectivity. However, the anatomical arrangement identified by Bassett et al. differs from ours, being primarily centred over the cortical systems relevant to the (sensorimotor) task. This suggests that the identity of cortical networks supporting slow, stable fluctuations may be somewhat context dependent, thus adapting according to prevailing requirements. From this perspective, it seems unlikely that cerebral cartography in its original incarnation will survive. It is possible that traditional brain mapping, as an isolated endeavour, will play a less prominent role in the future; the dynamic view entails an entirely different agenda that transcends what brain mapping traditionally had to offer. All, however, is not lost for cartography: Dynamic system theory brings a rich topographic language of its own, from the cartography of bifurcations to the complex space of attractors and their basins of attraction [e.g. (88)]. The future of cartography will thus likely witness a shift in focus from mapping brain anatomy and function, to more dynamic mappings reflecting complex, spatiotemporal neural dynamics that can be characterized with the application dynamic systems theory.

In our dynamical model, all regions are essentially identical - differing only by the total strength of external input. This in turn offsets the internal positive feedback through competitive action on local excitatory neurons (70). Intriguingly, recent research comparing macroscopic and local morphological connectivity suggests a variety of histological features that support this view, including more elaborate dendritic branching and larger soma size in high degree hubs, a histological counterpart of increased in-degree (61). Intriguingly, rich club regions show a variety of additional features such as a high density of dendritic spines (61) and high levels of energy consumption (89). Recent computational analyses have suggested that dendritic spines [as well as dendritic complexity (90)] slow down the passive conduction of dendritic currents between synapse and soma through subdiffusive anomalous "trapping" (91, 92). The effect of this is to stretch the relaxation time of the dendritic membrane, hence slowing down the dendritic response time (91). These empirical and modelling findings together suggest a slower membrane response time in hub pyramidal neurons. As a first cut, this can be accommodated in our generative model by reducing the time scale factor in the conductance model of our neural mass model – thus precisely mirroring the explicit model of time scale hierarchies introduced above (equations 1-2). In sum, just as the histological complexity of a neuronal population seem well matched to the complexity of their topological embedding (61), this histology also seems well matched to the time scales of their emergent dynamics. It remains to be seen whether these histological principles and associated dynamics occur in other species showing a rich club topology, from humans to the *C elegans*.

Recent empirical research has also focussed on the presence and function of time-dependent fluctuations in empirical recordings of functional connectivity, in EEG (54, 93, 94), MEG (95) and fMRI data (96-99); for review, see (100). The present analyses yield several testable predictions regarding the overlap between the time scales of temporal synchrony and the underlying connectome as well as the possibility of directly linking these changes to processes such as mood and introspection. Notably, the relationship between in-degree and neighbour synchrony clearly survived convolution with the hemodynamic response function (figure 3h) suggesting fMRI as a suitable approach in which to test our findings. It may also be that the presence of fast perceptual inputs (such as movie viewing) will further increase the rapid instabilities in peripheral (perceptual) regions (86). Spatially embedded temporal dynamics of cortical systems adds a rich layer of metrics for use in imaging research. Testing these predictions will, however, require judicious choice of several parameters, such as the appropriate bandpass of frequencies for the functional imaging modality, the state of awareness or arousal during which the resting-state fMRI data were acquired, and the appropriate resolution of the anatomical parcellation (101).

*3.3. Looking forward: Looking inward.*

We began with the premise that the time scales of neuronal activity should be matched to those of the "natural world" - hence the fast chattering of neuronal activity in systems supporting exteroception (the perception of external stimuli) and the slower time scales associated with internal states such as mood, motivation, and anxiety. Fast time scales – for example, the fleeting visual scenes whilst partaking in a 100m running race – are nested within the slower time scales; motivation for, and anxious anticipation of the race. This is a chronological take on the proposal that the complexity of cortical systems should match those of the world they represent (16). We end with a prospective that aims to tie this chronoarchitecture more formally to cognitive processes using the theory of predictive coding.

Predictive coding is a somewhat counter-intuitive account of cognition that dispels the naïve notion that we perceive the world through our senses: It states that instead we build internal models of the world to explain the causes of our sensations, and use action to test model-driven hypotheses (102, 103). Whilst relatively new to neuroscience (104), it has a rich and well established role in systems control engineering. Within a predictive coding perspective, the time scales of exteroceptive systems are necessarily fast as they endeavour to generate and update good predictive models of the (rapidly changing) external milieu (105). Consistent with the notion that "every good regulator of a system must be a good model of that system" (106), perceptual systems construct unstable models of rapidly changing sensory causes (86). Such models of the world then inform action, so that surprise (model violation) is minimised (40). Taking vision as an example, rapidly changing neuronal states - whose distributions encode (model) the likely states of the visual world - lead to fast eye movements (fixational eye movements and saccades) - that rapidly sample dynamic visual scenes (107). Put alternatively, the visual system - endowed with a high dynamic frustration - builds unstable perception "hypotheses" of the visual world (43), which then orchestrate fast fixational and saccadic eye movements (Figure 6, left hand side). In sum, dynamic frustration, and the ensuing itinerancy, endows brains with a rich and complex repertoire of representational or inferential states that can be used to select, through action, the next sensory inputs. This selection rests upon a generative model of the world that is as least as complex as the (local) environment being controlled. This is precisely Ross Ashby's law of requisite variety (108), which speaks to the notion that the complexity of cortical systems should match those of the world they represent.

We have proposed a novel "chronoarchitecture" of the brain, by which a slow dynamic cortical core emerges from the influence of the topological arrangement of the connectome on fast neuronal dynamics. This core comprises key cortical rich club regions such as the anterior insula, the anterior cingulate cortex (58) and the inferior frontal gyrus (109). Intriguingly, these regions are precisely those thought to support the perception of internal physiological states - namely interoception (110-112). Interoception has recently been positioned in the same predictive framework as exteroception (113): That is, internal (neural) models of likely physiological states are used to generate predicted changes in heart rate, perspiration and other autonomic percepts in response to (anticipated) emotionally salient stimuli (or internally generated material such as painful memories). Whereas motor behaviour (e.g. eye movement) underlies active inference in the exteroceptive system, active interoception involves anticipatory physiological changes mediated via the autonomic nervous system (114). Model violation leads to fluctuations in autonomic activity – such as skin conductance responses – and their associated emotional qualia (115). Interoception acts primarily to mediate inter-personal exchanges and hence unfolds on slower time scales than exteroception; whereas saccades have a time scale of ~3 Hz, heart rate variability has peak power at <0.1Hz and skin conductance responses occur several times a minute. Here we provide a computational account of the origin of the slower time scales in core, interoceptive systems (fig. 6, right hand side).

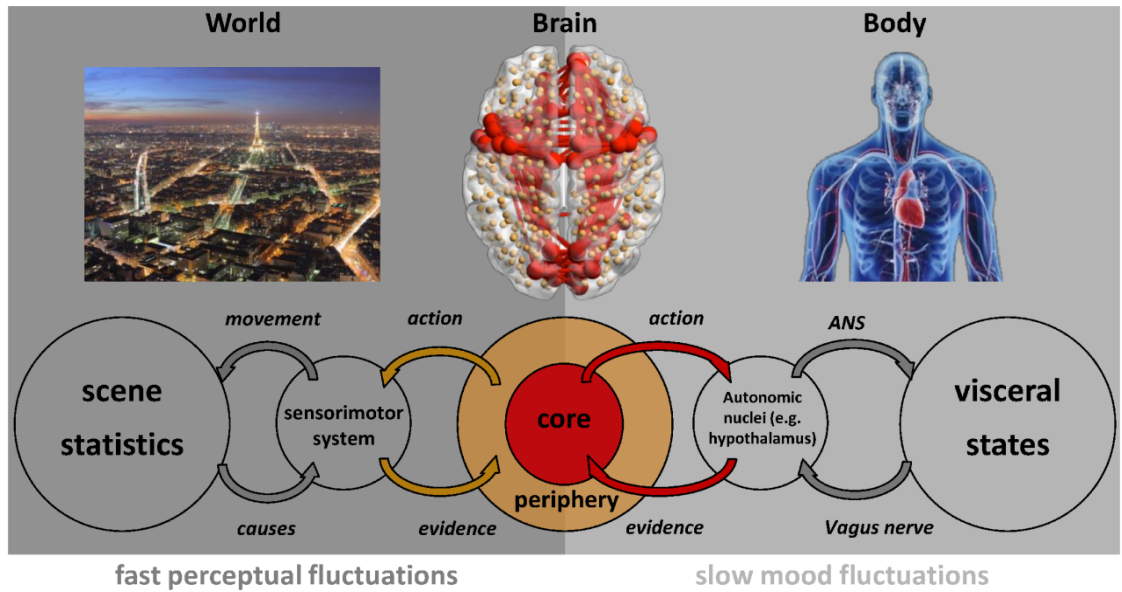

*Figure 6.* *The brain must cope with interconnected time scales: From fast perceptual to slow mood dynamics. Rapid exteroceptive perception is supported through fast, unstable dynamics in the topological periphery (left hand side). Slow interoceptive perception arises from slow neuronal dynamics within the topological rich club core of the brain (right hand side). Connectome image taken from reference (116).*

Disturbances in rich club topology have already been reported in mental illnesses such as schizophrenia (117). However, affective disorders may also be of greatest relevance to slow time scales of rich club regions. Melancholia and bipolar depression are associated with slowed mentation, longer sleeping and an impoverishment of thought (118, 119); in contrast, mania is associated with rapid speech, thinking and behaviour. Notably, rich club regions such as the inferior frontal gyrus and anterior insula have been broadly implicated in both major depression and bipolar disorder (120). Computational analyses of functional imaging data obtained from patients with melancholic depression shows key dynamical disturbances in network coupling and time scale parameters in the anterior insula, both at rest (121) and when viewing emotionally salient movies (122). Computational analyses of functional imaging data in young subjects at risk of bipolar disorder implicate dysfuncton of nonlinear influences amongst key rich-club regions such as the inferior frontal gyrus, anterior insula and dorsolateral inferior frontal gyrus (123). Further work is required to tie the "slow dysmetria" of affective disorders to the presently proposed substrates of slow neuronal dynamics: Our analyses may partly explain why many mental illnesses appear to involve cortical "hub-opathies" (124).

In sum, our account of cortical chronoarchitecture proposes a topological contribution – and a centre-periphery organization – to the slow time scales of the brain. However, the cortical core of these slow time scales does not support a disembodied introspection, but rather mediate active inference of bodily states. This dispels the notion of "pure reason" (125), instead situating core cortical processes in a homeostatic relationship with the body.

**Acknowledgements**

The authors thank Alistair Perry for the use of the connectome image in Figure 6.

**Funding Statement**

LLG and MB acknowledge Australian Research Council (Centre of Excellence in Integrative Brain Function, CIBF) and the NHMRC (Program Grant APP 1037196). RMH acknowledges the Canadian Institute of Health Research (CIHR) postdoctoral fellowship and Brain and Behaviour Research Foundation NARSAD Young Investigator Grant. AZ is supported by the NHMRC Career Development Fellowship (GNT1047648).

# Appendix A

*A1. Dynamics*

Representing cortical regions, nodes were modelled by conductance-based neural mass models (70). This simple model with rich dynamics has proved to be particularly suitable to simulate large-scale cortical networks (41, 66, 67, 99, 126). Each cortical region *i* receives input from its $K_{in}^i$ afferent regions, where $K_{in}^i$ is the in degree of node *i*. Its dynamics is described by three nonlinear differential equations:

$$\frac{dV^i(t)}{dt} = -\left\{g_{Ca} + r_{NMDA}\, a_{ee}\left[(1-c)Q_V^i + c\langle Q_V^j(t-\tau)\rangle\right]\right\} m_{Ca}\left(V^i(t) - V_{Ca}\right)$$

$$- \left\{g_{Na} m_{Na} + a_{ee}\left[(1-c)Q_V^i + c\langle Q_V^j(t-\tau)\rangle\right]\right\}\left(V^i(t) - V_{Na}\right)$$

$$- g_K W^i(t)\left(V^i(t) - V_K\right) - g_L\left(V^i(t) - V_L\right) - a_{ie} Z^i(t) Q_Z^i + a_{ne} I_\delta ,$$

$$\frac{dZ^i(t)}{dt} = b\left[a_{ni} I_\delta + a_{ei} V(t) + Q_V^i(t)\right] ,$$

$$\frac{dW^i(t)}{dt} = \frac{\varphi\,[m_k - W^i(t)]}{\tau_W} .$$

Where V represents the mean membrane potential of the excitatory pyramidal neurons, Z the mean membrane potential of the inhibitory interneurons, and W the average number of potassium ion channels. And the network contribution is:

$$\langle Q_V^j(t-\tau)\rangle = \frac{1}{K_{in}^i}\sum_{j=1}^{K_{in}^i} Q_V^j(t-\tau) .$$

The fraction of open channels is given by

$$m_{ion} = 0.5\left[1 + \tanh\left(\frac{V^i(t) - T_{ion}}{\delta_{ion}}\right)\right] .$$

And the average firing rate of excitatory and inhibitory populations are described by the respective sigmoidal activation functions,

$$Q_V^i = 0.5\left[1 + \tanh\left(\frac{V^i(t)}{\delta_V}\right)\right],$$

$$Q_Z^i = 0.5\left[1 + \tanh\left(\frac{Z^i(t)}{\delta_Z}\right)\right].$$

Finally, the parameters used were: $g_{Ca}$=1.1, $r_{NMDA}$=0.25, $a_{ee}$=0.4, $V_{Ca}$=1, $g_{Na}$=6.7, $V_{Na}$=0.53, $g_K$=2, $V_K$=-0.7, $g_L$=0.5, $V_L$=-0.5, $a_{ie}$=2, $a_{ne}$=1, $I_\delta$=0.3, b=0.1, $a_{ni}$=0.4, $a_{ei}$=2, $T_{Ca}$=-0.01, $T_{Na}$=0.3, $T_K$=0, $\delta_{Ca}$=0.15, $\delta_{Na}$=0.15, $\delta_K$=0.3, $\varphi$=0.7, $\tau_W$=1, $\delta_V$=0.65, and $\delta_Z$=0.65. In the results presented we fixed the delay $\tau$=15 ms (or to $\tau$=0 ms for the sake of comparison), and the coupling strength c=0.01 for consistency. Our results are, however, robust for other values of coupling strength and conduction delays.

*A2. Simulation details*

Simulations were performed in Matlab using the function dde23. Except for short illustrative time traces and long simulations of 600 s (that were convolved with the HRF), results represent an average of 20 trials of 90 s after a transient period of 600 ms, which is discarded.

*A3. Normalization of node neighbours.*

Since neighbour synchronization is a stochastic measure, we need to average over several trials to obtain a reliable estimative. First we averaged over ten trials the difference between the crosscorrelation of each pair of nodes' neighbours and the crosscorrelation between a pair of randomly selected nodes. Then, averaging this over all possible pairs of each node's neighbours we have the neighbour synchronization.